\documentclass[12pt]{article}

\usepackage{amssymb, amsmath, amscd}
\usepackage{graphicx}
\usepackage[latin1]{inputenc}
\usepackage{epsfig}
\usepackage{color}
\usepackage{url}

\usepackage{authblk}

\usepackage{algorithm}
\usepackage{algorithmic} 
\usepackage[table]{xcolor}
\usepackage{tikz-cd}
\newtheorem{definition}{Definition}

\numberwithin{equation}{section}

\usepackage{pdflscape}
\usepackage{afterpage}
\usepackage{capt-of}
\begin{document}

\bibliographystyle{abbrv}

%\title{Quantum topological de novo discovery of mutated driver pathways in cancer}
\title{The Topology of Mutated Driver Pathways}
%\title{{\sc Topological approach to cancer genomics pathways}
\author[1]{Raouf Dridi\footnote{rdridi@andrew.cmu.edu}}

\author[1]{Hedayat Alghassi\footnote{halghassi@cmu.edu}}
\author[2]{Maen Obeidat\footnote{Maen.Obeidat@hli.ubc.ca}}
\author[1]{Sridhar Tayur\footnote{stayur@cmu.edu}}

\affil[1]{Quantum Computing Group, Tepper School of Business,
Carnegie Mellon University, Pittsburgh, PA}
  
\affil[2]{The University of British Columbia Center for Heart Lung Innovation, St
Paul's Hospital Vancouver, Vancouver, BC}

\maketitle

\begin{abstract}
Much progress has been made, and continues to be made, towards identifying candidate mutated driver pathways in cancer. However, no systematic approach to understanding how candidate pathways relate to each other for a given cancer (such as Acute myeloid leukemia), and how one type of cancer may be similar or different from another with regard to their respective pathways (Acute myeloid leukemia vs. Glioblastoma multiforme  for instance), has emerged thus far. 
	 Our work attempts to contribute to the understanding of {\em space of pathways} through  a novel topological framework.
 We illustrate our approach, using  mutation data (obtained from TCGA) of two types of tumors:  Acute myeloid leukemia (AML) and Glioblastoma multiforme (GBM). We find that the space of pathways for AML is homotopy equivalent to a sphere, while that of GBM is equivalent to a genus-2 surface.  We hope to trigger new types of questions  (i.e., allow for novel kinds of hypotheses) towards a more comprehensive grasp of cancer.
	 
	 %As a first application, we  explain how the noisy data problem in the data-driven pathways discovery is dealt with within this framework using  persistent homology. 
	   %We present a complete persistent homology-based algorithm built on the top of our quantum procedure.  For both datasets, using our  homology based  quantum pipeline, we have recovered the same pathways previously found in addition to new additional pathways. 

%	   {\it ``How can it be that mathematics, being after all a product of human thought which is independent of experience, is so admirably appropriate to the objects of reality?" [Albert Einstein]}
%
%	{\it ``We should stop acting as if our goal is to author extremely elegant theories, and instead embrace complexity and make use of the best ally we have: the unreasonable effectiveness of data." [Peter Norvig]}
	
\end{abstract}	  

%{\color{blue} Cite "Topology and data" and "Ayasdi work in breast cancer" }
%	 ~~\\
	 
	 	 ~~

 \section{Introduction}
  Let us begin by recalling a quote of Henri Poincare:
 \begin{quote}
     {Science is built up of facts, as a house is built of stones; but an accumulation of facts is no more a science than a heap of stones is a house.}
 \end{quote}
Cancer is driven by somatic mutations that target signaling and regulatory pathways that control cellular proliferation and cell death~\cite{bioOfCancer}. Understanding how this happens is of paramount importance 
in order to improve our ability to intervene and attack cancer. Since  the advent  of DNA sequencing technologies, our understanding has progressed enormously and resulted in useful therapies.\footnote{Current advances include the introduction of a single anti-cancer agents, which simply bind with growth factor receptors stopping abnormal cell 
proliferations. For instance, in the context of  breast cancer, 
  Herceptin antibody stops cells abnormal proliferation signals by binding  with the excess of growth factor receptors on the cell surface  caused by point mutations in the gene Her2.   Gleevec  is a second example. It is used to treat  chronic myeloid leukemia, which is a type of blood cell tumor due to an inappropriate gene fusion product
of a translocation in chromosomes 9 and 22. The resulting fusion gene BCR-ABL has an increased kinase activity, resulting in an increase in 
proliferation signals.  Similar  to Herceptin, Gleevec blocks the growth signals that the abnormal fusion gene generates and thus prevents the cell proliferation. More sophisticated approaches include  immunotherapy that activates  the immune system against cancer cells as well as the use of combinations of therapeutic agents to
attack multiple pathways fundamental in cancer development, preventing resistance from occurring.}  %But, much more work remains to be done.

%For example, recent researches have led to the ability to activate the immune system against cancer cells.  Called immunotherapy, this approach has shown success with melanoma, leukemia, and lymphoma, and is ripe for further exploration in a wider range of cancers.  Another approach attacks multiple pathways fundamental in cancer development, using combinations of therapeutic agents to prevent resistance from occurring.  %But, much more work remains to be done.

%This led to  the introduction of very efficient molecularly target anti-cancer agents. 
~~\\
Notwithstanding the above, cancer morbidity is still very high and
 our understanding is still incomplete.  Certainly, finding more relevant pathways will be helpful. Indeed we have developed novel formulations and algorithms that are computationally effective in doing so, a subject of our companion paper (\cite{alghassi_quantum_2019} that builds on our work \cite{Hedayat2, Hedayat1}). We also believe that to start building a house it is  not enough to simply accumulate more stones faster. To that end, in this paper, we look at whether the stones that we have identified so far form patterns that can help us build a house, using methods from algebraic topology (see Table 1).
 
\begin{figure}[h]\label{exples}
	\begin{center}
      \includegraphics[scale=0.35]{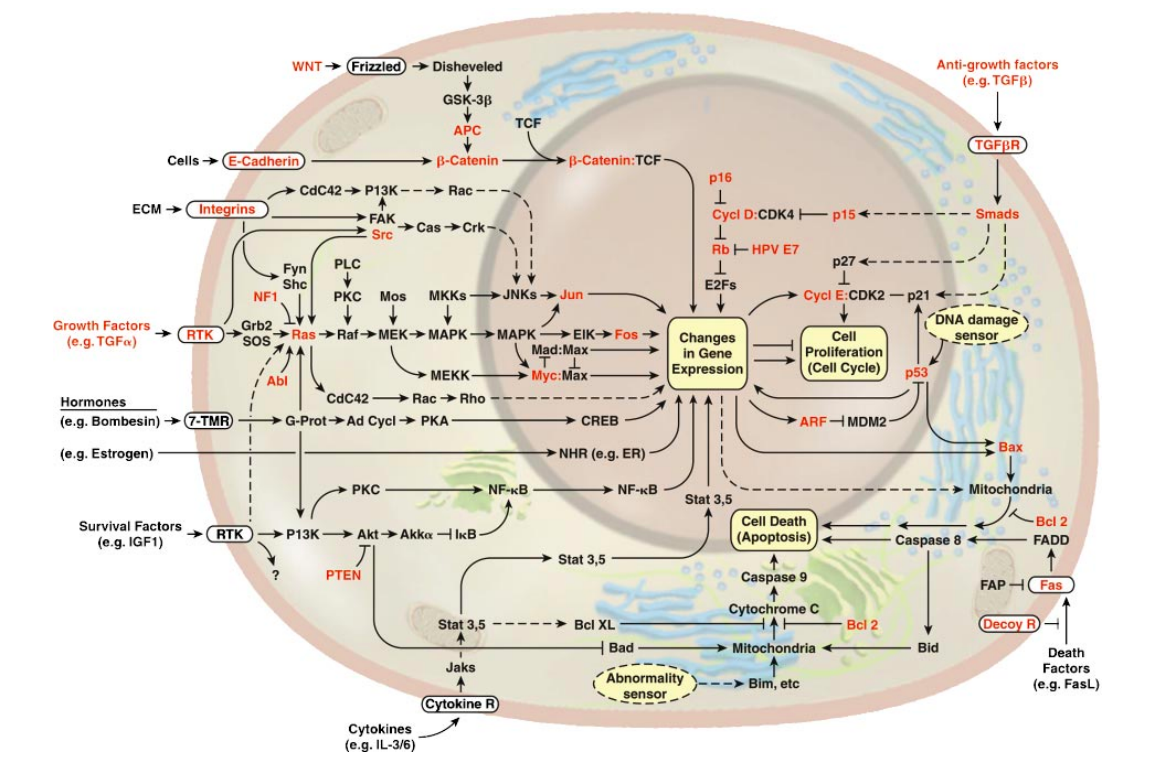} 
      \end{center}
    \caption{Cell circuitry,  wiring pathways, resembles electronic integrated circuits. A full understanding of this picture will, one day, replace the present day patchwork with a more rational approach based on conceptual  structures and logical  coherence similar to those encountered in  chemistry and physics~\cite{hallmarksOfCancer}. }
\label{circuit}
\end{figure}

 ~~\\
 The view we are taking here is also articulated in {\em The Hallmarks of Cancer}~(\cite{hallmarksOfCancer}):
 \begin{quote}
     Two decades from now, having fully charted the wiring diagrams
     of every cellular signaling pathway, it will be possible  to lay out the complete ``integrated circuit of the
    cell" upon its current outline ({Figure~\ref{circuit}}). We will then be able to apply the tools of  mathematical modeling  to explain how specific genetic lesions serve to reprogram this integrated circuit... One day, we imagine that cancer biology and treatment--at present, a patchwork quilt of cell biology, genetics, histopathology, biochemistry, immunology, and pharmacology--will become a science  with a conceptual structure and logical coherence that rivals that of chemistry or physics. 
 \end{quote}

\begin{table}[h]
\centering
\begin{tabular}{|c|c|}
\hline
Pathways & Algebraic topology \\
\hline \hline
Regulatory/signaling pathway & simplex \\
The Emergent integrated circuit of the cell & simplicial complex $K_\varepsilon$\\
Number of patients shared between two genes & persistent parameter $\varepsilon$\\
%\hline\\
%Optimal implementation of $U$ &  a geodesic with $U(0)=I$ and $U(t_f)=U$  \\
\hline
\end{tabular}
\caption{Correspondence between pathways and algebraic topology.}
\end{table}

~~\\ 
Our topological journey departs from the (well trodden) path that recognized that signaling or regulatory  pathways can be viewed as {\em independent sets} (modulo 
some notion of tolerance) in a matrix with rows as patients and columns as genes,  which is used in determining new pathways via different computational methods (see our companion paper \cite{alghassi_quantum_2019} and previous works \cite{Vandin, Vandin2, Ciriello:2012:Genome-Res:21908773,Szczurek, Leiserson}).
Our key insight is that these pathways (however discovered), when grouped together,  define  a {\it simplicial complex} which is, pictorially, a polytope  with faces given by  those pathways. 
%collection of objects called {\it faces} closed under some boundary map  i.e., the boundary of a face is again a face. 
% A clique (resp. independence)  complex of a graph $G$ is a simplicial complex whose faces 
%are cliques  (resp. independent sets)  of $G$ and boundary map is the inclusion of sets). 
This vantage point connects us to the marvellous world of topology where simplicial complexes are the prototypes of  spaces with shapes. Our journey explores this {\em notion of shape} in cancer genomics.  

~~\\
Our framework is built on this sequence of assignments: 

 ~~\\ 
  \begin{tabular}{ccccccc}\label{functor3}
	{\bf MutationData} & 		$\rightarrow$      & {\bf graphs} & $\rightarrow$& {\bf SpaceOfPathways}  		& $\rightarrow$& {\bf ShapeMeasurements} \\\nonumber
	$X$ 			   &           $\mapsto$ 	 &   $G$            &$\mapsto$  &    $ \mathcal N(K)$                             &$\mapsto$ &      $H_*(\mathcal N(K))$\\
\end{tabular}
%
%
%\begin{eqnarray}\label{functor3}
%	{\bf mutationData} & 		\rightarrow& 		{\bf graphs} \rightarrow {\bf simplicialComplexes}  			j\\\nonumber
%	X 			   &\mapsto 		&              \quad G \quad   \quad \mapsto  \mathcal N(K)  \quad \qquad \mapsto \\
%\end{eqnarray} 

~\\
where: 
\begin{itemize}
\item $G$ is the mutation graph. That is, 
is the graph defined by the set of genes in $X$  where two genes are connected if they have harboured mutations for the same patients.
  $K$ is the simplicial complex  given by the set of all independent sets of $G$.  %(It is important to mention again that the construction of the simplificial complex $K$ is prohibitively hard task for classical computers. This is where our algorithms
% developed elsewhere are used). 
 The underlying mathematics detailed in \cite{raoufHomology} is based on the so-called Mayer-Vietoris construction, which itself is articulated around clique covering the graph $G$. We prove in \cite{raoufHomology}  that, for a large class of graphs,  such coverings provide most of the simplices of $K$.  %Another quantum approach which can be used here as well is the quantum algorithm presented in \cite{seth} written in the gate model (our is written as
%an adiabatic quantum computation). 
\item $  \mathcal N(K)$ is the space of pathways. The facets (maximal independent sets) of $K$ are the pathways, and their nerve defines the {space of pathways} $  \mathcal N(K)$. Pictorially, 
the space of pathways is visualized through its 1-skeleton of the graph with pathways as vertices and where two pathways
are connected by an edge if they intersect. 
\item $H_*(\mathcal N(K))$  is the homology of  the space of pathways, that is, the shape measurements of the space of pathways. 
\end{itemize}
	 We have applied our approach to two different  
	 mutation data obtained from TCGA: Acute myeloid leukemia (AML) \cite{doi:10.1056/NEJMoa1301689} and Glioblastoma multiforme (GBM) \cite{GBM2008}. For both the data,
	 we have computed the assignment 
	 \begin{equation}
	     \mbox{tumor} \rightarrow \mbox{space of pathways}
	 \end{equation} 
	 using persistent homology.\footnote{In the language of {\it Homotopy Type Theory (HoTT)} \cite{hottbook}, this procedure translates into: {{\it tumor $\mapsto$ 
	 $(\infty, 1)-$category}}. This $(\infty, 1)-$category is an $(\infty, 1)-$topos (a logical structure)  in the sense of Lurie \cite{10.2307/j.ctt7s47v}, so it obeys the {\it axioms} of HoTT.
	 }  Our calculation  shows  that the space of pathways for 
	AML mutation data is homotopy equivalent to a sphere, while in the case of GBM data, the space of pathways is homotopy equivalent to figure eight (genus-2 surface).
	\\~~
	
	 %Computations here are performed using D-Wave 2X quantum computer. 

  %\cite{ayasdi, Nicolau2011TopologyBD, Yao2009TopologicalMF, CoMet, CoMetNature, CoMetNatureCom, CoMetNature2019, NoMas, VANDEHAAR20191375} . 
 
 %\begin{figure}[h]\label{flowchart}
%	\begin{center}
%      \includegraphics[scale=0.5]{pics/QMLflowchart.png} 
%      \end{center}
%    \caption{Flowchart of the proposed procedure for driver pathways determination and space of pathways construction. The parameter $\varepsilon$ is the number of patients shared between mutated genes. It takes a finite number of increasing values.}
%\label{circuit}
%\end{figure}

 \section{Related works}
 The  motivation for the present work  originates  from the work of the Raphael Lab, centred around the Dendrix algorithm \cite{Vandin01022012},  and its later improvements including CoMEt \cite{CoMet}.  Both algorithms  are in widespread use in  whole-genome analysis--for instance, in \cite{CoMetNature, CoMetNatureCom, CoMetNature2019,  VANDEHAAR20191375}.
 Building on those foundations, our work  extends   in the following two directions: First, the two key notions of exclusivity and coverage are abstracted here by the two simplicial complexes $K_\varepsilon$ and $K_\eta$ or, more precisely, by the filtrations of simplicial complexes $ K_{\varepsilon_0} \subset \cdots \subset K_{\varepsilon_\ell} $ and $ K_{\eta_0} \subset \cdots \subset K_{\eta_{\ell'}} $,  where we allow the  two parameters to vary:  $\varepsilon_0\leq\varepsilon\leq \varepsilon_\ell$ and $\eta_0\leq\eta\leq \eta_{\ell'}$. 
 We do so  because, in reality,  the assumptions that driver pathways exhibit both high coverage and high exclusivity need not to be strictly satisfied.  The functoriality of persistent homology, which takes the two filterations as input, handles this   elegantly; we decide on the values of these two parameters after computing the barcodes.
 This type of analysis is not present in the aforementioned works.  
 Second,  
 motivated by the {\it naturality} of the constructions, the present paper goes beyond the computational aspects  and ventures into the {\it conceptuality of cancer}.  We have introduced  the notion of the topological space of pathways  $\mathcal N$, together with its homology spaces,  as a  paradigm to rationalize the extraordinary complexity of cancer.  To the best of our knowledge, this is a   ``provocative" idea that has not been  explored before.

~~\\ 
 Another related set of works  is \cite{ayasdi} and \cite{Nicolau2011TopologyBD} (from Stanford's applied topology group), which 
 also use algebraic topology tools in cancer. 
 However, they differ from ours on two counts:  the nature of the problem treated and the  methodology used. The algorithm introduced, called Mapper, is  a topological clustering algorithm, and it is not based on persistent homology. Mapper was used to identify a subgroup of breast cancers with  excellent survival, solely based on topological properties of the data.\footnote{Commercial applications of this approach in various different areas of practice are tackled through Ayasdi.}

%\url{https://bit.ly/2PxfRJg}
  
% {\color{blue}
% We   position our work vis-\`a-vis 
%  related literature. 
 %, where advanced mathematical approaches in cancer research are  central
 %--including graph theory-based driver pathways determination and topological clustering of gene expression data-- 
% \begin{itemize}
%     \item  The Mapper method (from Stanford's applied topology group)
%   \cite{ayasdi}  is a topological clustering method that is used  to cluster gene expression data. As an example, in \cite{Nicolau2011TopologyBD}, Mapper identified a subgroup of breast cancers with a unique mutational profile and excellent survival, solely based on topological properties of the data.\footnote{Commercial applications of this approach in various different areas of practice are tackled through Ayasdi.}
 %  \item The     
% Dendrix algorithm \cite{Vandin01022012} (from Raphael Lab), and its later improvements including CoMEt \cite{CoMet}, are procedures for the discovery of mutated driver pathways in cancer using only mutation data (which is the same here and in \cite{alghassi_quantum_2019}).  Dendrix  and CoMEt  are in widespread use in  whole-genome analysis--for instance, in \cite{CoMetNature, CoMetNatureCom, CoMetNature2019,  VANDEHAAR20191375}. 
% \end{itemize}
%Building on those foundations,  our paper investigates the space of pathways, computationally {\it and} conceptually, using {\it  persistent homology}.  To the best of our knowledge, this is a novel idea that has not been  explored before.   }

 \section{Connecting to algebraic topology}
Different errors occurring during data preparation (i.e., sequencing step, etc.) affect the robustness of the results. This implies that the computed pathways are likely to be affected by these errors  and can not be considered as a robust finding without explicitly modeling the error in our constructions. 
To that end, the assignment we mentioned above is done as follows:   
 \begin{itemize}
\item [1.] We think about the number of patients that are shared between two genes
 as a parameter $\varepsilon$ (thus, absolute exclusivity corresponds to taking this parameter to zero). 
 \item  [2.] Instead of applying our procedure once, we apply it for a range of values
 of the exclusivity parameter $\varepsilon$. That is, we consider  a filtration of graphs (instead of one): 

	  \begin{equation}\label{firstFiltration}
	  	G_{\varepsilon_0} \subset \cdots \subset G_{\varepsilon_\ell} 
	\end{equation}
	  where for each graph  $G_{\varepsilon}$ 	
	 two genes are connected if they have harboured mutations  concurrently for at least $\varepsilon$ patients. This yields a second
	 filtration of simplicial complexes
	  (we call  such a filtration, a {\it persistent pathway complex}) 
	\begin{equation}\label{secondFiltration}
	 K_{\varepsilon_0} \subset \cdots \subset K_{\varepsilon_\ell} 
	 \end{equation}
	  where $K_{\varepsilon}$ is one of the three complexes we define below.  
  \item [3.] Measure the shape (the homology) of the different pathway spaces and then ``average"  the shape measurements that are obtained.  
 \end{itemize}
 This (practical) version of homology is what we refer to as persistent homology. It tracks the persistent topological features through
a range of values of the parameter; genuine  topological properties  persist through the change of the parameter whilst noisy observations do not (all of these will be made precise below). The key point is that the mapping $G_\varepsilon \mapsto K_\varepsilon$ is  {\it functorial}; that is, it sends
 a whole filtration (i.e., \ref{firstFiltration}) into another filtration (i.e., \ref{secondFiltration}). In other words, it is not only sending graphs to simplicial complexes but it is also preserving their 
 relations.
 %\footnote{In fact, we have a sequence of functors $(\mathbb R^+, \leq) \rightarrow   ( {\bf graphs}, \subseteq)  \rightarrow   ( {\bf simpComplexes}, \subseteq) $
% which sends $\varepsilon$ to $G_\varepsilon$ to $K_\varepsilon$.  
 %}. 
This functoriality is at the heart of persistent homology. 

 \subsection{The Gene-Patient graph} 
Consider a mutation data for $m$ tumors (i.e., patients), where each of the $n$ genes is tested for a somatic
 mutation in each patient.  To this data we associate a mutation matrix $B$ with $m$
	rows and $n$ columns, where each row represents a patient and each column
	represents a gene. The entry $B_{ig}$ in row $i$ and column $g$ is equal to 1 if patient $i$ harbours a mutation in gene $g$ and it is 0 otherwise.  
For a gene $g$,   we define the fiber
\begin{equation}
	{\sf Patients}(g) \mbox { = the set of patients in which $g$ has
	 mutated.  }
\end{equation}	 
 \begin{definition}
	
	The \emph{mutation
	graph} associated to $B$ and $\varepsilon>0$ is the graph $G_\varepsilon$  whose vertex set is the 
	set of genes and whose edges are pairs of genes $(g,\, g')$ such that
	\begin{equation}
	{|{\sf Patients}(g)\cap{\sf Patients}(g')|\geq \varepsilon.}
	\end{equation}
\end{definition}
	{There are evidences that pathways can be treated as independent sets of mutation graphs (although not stated graph-theoretically) \cite{Vogelstein, Yeang01082008}. }

 \subsection{The space of pathways} \label{complexDefs}
We would like to assign to the mutation graph  an independence complex. We present below two different functorial ways to do so. 

 \begin{definition}
 Given a mutation graph $G_\varepsilon$, its \emph{persistent pathway complex}~$K_\varepsilon$ is the
 independence complex of $G_\varepsilon$ (or equivalently, the clique complex of~$\overline G_\varepsilon$, the complement of $G_\varepsilon$). 
  \end{definition}
  We also define the persistent pathway complex $K_\eta$. 
  \begin{definition} The  persistent pathway
  complex $K_\eta$ is defined as follows. Fix  $\varepsilon=\varepsilon_0$ and let $G=G_{\varepsilon_0}$. 
  The complex  $K_\eta$ is the complex generated  by all independent sets $S$ of $G$ with coverage
 \begin{equation}
 	\sum_{g\in S} {\sf Patients}(g) \geq \eta
\end{equation} where the counting ${g\in S}$ is done without redundancy. 
    \end{definition}
%We note here that the way the complex $K_\alpha$ is constructed is different then the assignment  (\ref{functor3}) as for $K_\varepsilon$  and $K_\eta$. The complex $K_\alpha$ is constructed by intersecting all pathways computed using \cite{alghassi_quantum_2019}. 
% {\color{blue} Functoriality of $K_\alpha$}
 The following definition is also valid for the persistent pathway complex $K_\eta$.  
  \begin{definition} The space of pathways of a persistent pathway complex $K_\varepsilon$  is  the nerve generated by the facets
  of  the complex, that is, the simplicial complex where $\{i_0, \cdots, i_\ell\}$ is a simplex if and only if 
  the facets indexed with $i_0, \cdots, i_\ell$ have a non empty intersection. We denote the space of pathways by $\mathcal N_\varepsilon$. 
    \end{definition}
The space of pathways is visualized through its 1-skeleton: the graph with pathways as vertices, and two pathways
are connected if they intersect (see Figures 3 and 4).  

%~~\\
%The construction of the independence complex consists of enumerating all independent sets in the given graph. This is clearly a very expansive computation,  for which  we propose the use of quantum computing.
%We have also developed hybrid quantum-classical and fully classical exact algorithms for finding the optimal covering  \cite{Hedayat2, Hedayat1}, which is the key ingredient in persistent homology computations.
%We note the existence of   classical heuristics  which are available for small graphs. For instance,
%Bron and Kerbosch's algorithm \cite{Bron:1973:AFC:362342.362367, CAZALS2008564} is used in {\sf Python} graph library {\sf networkx}.  
%This algorithm  runs out of memory for large graphs. 

 \subsection{A primer on persistent homology} 
Recall that  our plan is to compute the persistent homology of the independence complex $K_\varepsilon$. We have explained that this is simply computing the homology of $K_\varepsilon$
for a range of increasing values of the parameter $\varepsilon$. In this section, we explain this notion of homology (which we have introduced as measurements of the shape of the space $K_\varepsilon$). For simplicity, we drop out  the subscript $\varepsilon$ from the complex $K_\varepsilon$. 
It is also more  convenient to introduce homology  for clique complexes (for independence complexes, it suffices  to replace, everywhere below, cliques
with independent sets).

\begin{figure}[!htbp]
\centering
\begin{minipage}{1.0\textwidth}
\centering
\includegraphics[width=0.3\linewidth, height=0.15\textheight]{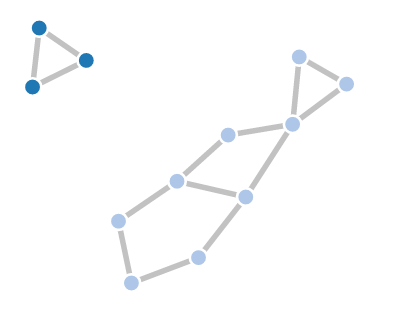}
\caption{The  graph has two connected components, which implies $\beta_0=2$. It also has two cycles that are not triangles; thus, $\beta_1=2$. Higher Betti numbers are zero. }
\label{h01}
\end{minipage}
\begin{minipage}{1.0\textwidth}
\centering
\includegraphics[width=0.4\linewidth, height=0.15\textheight]{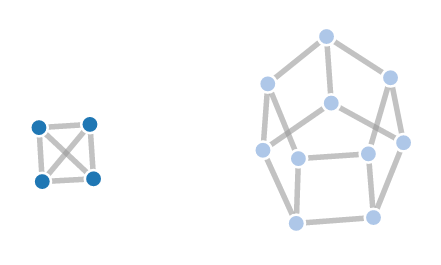}
\caption{The  graph has two connected components, giving $\beta_0=2$. It also has 7 cycles that are not  triangles, which yields $\beta_1=7$. Higher Betti numbers are zero here as well. Now, if the different sides of the hexagonal prism (right component) are covered with
    triangles, then we get instead $\beta_1=0$ and  $\beta_2=1$.}
\label{h2}
\end{minipage}
\end{figure}

~~\\
 The homology of the simplicial complex $K$ (now a clique complex of $G$) 
 is a sequence of  $\mathbb Z$-vector spaces (i.e., vector spaces with integer coefficients): 
\begin{equation}
	H_*(K) := H_0(K), \quad H_1(K), \quad H_2(K), \quad \dots
\end{equation}
defined as follows:  
\begin{itemize}
    \item The  zeroth vector space~$H_0(K)$ is spanned by all connected components of~$K$; thus, the dimension $\beta_0:= \mathrm{dim} (H_0(K))$ gives
the number of connected components of the space. 

\item The first homology space~$H_1(K)$ is spanned by all  closed chains of edges (cycles) in $G$ which are not triangles -- see Figure \ref{h01}; in this case, 
the dimension $\beta_1:= \mathrm{dim} (H_1(K))$ gives
the number of ``holes" in the space.

\item Similarly, 
the second space
$H_2(K)$ is spanned by all 2-dimensional enclosed three dimensional ``voids"  that are not  tetrahedra (as  in Figure~\ref{h2} below).  
\end{itemize}
Higher dimensional spaces
are defined in a similar way (although less visual).  Their dimensions count non trivial high dimensional voids. 
  The dimensions $\beta_i := \mathrm{dim } (H_i(K))$ are called Betti numbers and provide the formal description of the concept of shape measurements.\footnote{In textbooks,  one typically starts with a continuous space (for instance, a doughnut shaped surface) and then triangulizes it, yielding the simplicial complex $K$.  The homology of the continuous space is the homology of its discretization $K$ that we introduced pictorially above. Appendix \ref{SectionAppendix}, on page~\pageref{SectionAppendix}, explains how to compute homology using quantum computers or  QUBO-solvers such as \cite{Hedayat2}.}

~~\\
We move now to the notion of  persistent homology and make it a bit more precise. For that, let us reintroduce the persistent parameter $\varepsilon$ and let $K_\varepsilon$ be again an independence complex.   It is clear that  if 
$|{\sf Patients}(g)\cap{\sf Patients}(g')|\geq \varepsilon_1  $ and $ \varepsilon_1 \geq \varepsilon_2$
then the pair $(g, g')$, which is an edge in ${G}_{\varepsilon_1},$ is also is an edge in $G_{\varepsilon_2}$. This means that
${G}_{\varepsilon_1}$ is a subgraph of ${G}_{\varepsilon_2}$; thus, we have
  $K_{\varepsilon_2}\subset  K_{\varepsilon_1}$ whenever $\varepsilon_1 \geq \varepsilon_2$ (since an independent set for a given graph is also an independent set for any of its subgraphs). The mapping $\varepsilon \mapsto K_\varepsilon$ is functorial.
It turns out that homology itself is  functorial  and
all  this functoriality  is the mathematical reason that the following is correct: one can track the Betti numbers over a range 
of values ${\varepsilon_1 \geq \varepsilon_2 \geq \varepsilon_3\geq \cdots}$ and consider the subrange where 
the Betti numbers are not changing (significantly). Pathways within this subrange are considered to have passed our test and declared robust computations.

\section{Real mutation data}
We have applied our approach to two mutation data (formulations and algorithms are available in \cite{alghassi_quantum_2019}):  Acute myeloid leukemia \cite{doi:10.1056/NEJMoa1301689} and Glioblastoma multiforme \cite{GBM2008}. For both data, we have computed the assignment~{\it tumor $\mapsto$ pathways} through persistent pathway complexes (thus, declared robust output).  
The complete result is presented in long tables given in the Appendix. 
Interestingly, our calculation also shows  that 
AML data is homotopy equivalent to a sphere while GBM data is homotopy equivalent to figure eight (genus-2 surface). 

\subsection{Acute myeloid leukemia data}
The data has a cohort of 200 patients and 33 genes (\cite{doi:10.1056/NEJMoa1301689}).  We have chosen the coverage threshold $\eta=80$ patient. We also neglected all genes 
that have fewer than 6 patients.  These numbers are chosen  based on the stability of barecodes for pairs $(\varepsilon, \eta) \geq (6, 80)$, while
barecodes for pairs less than (6, 80) exhibit strong variations.  This is also consistent with the fact that choosing genes   with fewer than  5 or 6  patients is  not common in such studies (genes with low numbers of patients are not considered robust enough, and are very prone to errors. This   extra precaution is commonly used in the field).    
Now for the numbers of patients and coverage we have chosen, the Betti numbers are computed for various values of $\varepsilon$ in the table below:
\begin{center}
\begin{tabular}{cccc}
	$\varepsilon$ &   $|\mathcal N_\varepsilon|$ &  $density(\mathcal N_\varepsilon)$  & $\beta_i$\\
	\hline\\
	                   1 &               					   6 & 								0.86	& $1, 0,0, \cdots$\\
	\hline\\
				2 &							84	& 						0.97		&	 $1, 0,0, \cdots$ \\
	\hline\\
				3 &							50	& 						1		&	 $1, 0,0, \cdots$ \\
\end{tabular}
\end{center}
Figure \ref{AMLnerve} below gives the 1-skeleton of the nerve $\mathcal N_\varepsilon := \mathcal N(K_\varepsilon)$ for $\varepsilon=1$. The Betti numbers $\beta_i$ are not changing;  thus, $\varepsilon=1$ is a reasonable choice. 
Recall that each  node represents a pathway and two pathways are connected if they intersect (as sets of genes). We have used different colors to represent different pathways as described in the table below (no other meaning for the coloring).  

\begin{figure}[h]\label{exples}
	\begin{center}
      \includegraphics[scale=0.5]{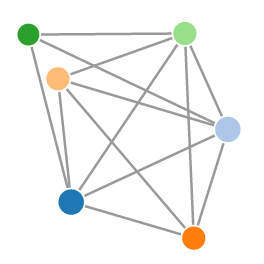} 
      \end{center}
    \caption{{\small The 1-skeleton of the nerve $\mathcal N_\varepsilon$ for $\varepsilon=1$ for AML data. 
    In this case, the nerve is homotopy equivalent to the sphere. The different pathways represented by the nodes are given in the table above.}
    %The pathways here have less chance
    %to be affected by the noise. 
}
\label{AMLnerve}
\end{figure}

~~\\
{\footnotesize
\begin{tabular}{l|l}
 \hline\\
 Color & Genes in the pathway\\
 \hline\\
 Blue &  'PML.RARA', 'MYH11.CBFB', 'RUNX1.RUNX1T1', 'TP53', 'NPM1', 'RUNX1' \\
  \hline\\
 Blue light & 'PML.RARA', 'MYH11.CBFB', 'RUNX1.RUNX1T1', 'TP53', 'NPM1', 'MLL.PTD'\\	
  \hline\\
  Orange &  'PML.RARA', 'MYH11.CBFB', 'RUNX1.RUNX1T1', 'DNMT3A'\\
   \hline\\
  Orange light & 'Other Tyr kinases', 'MYH11.CBFB', 'MLL.PTD', 'NPM1'\\
   \hline\\
      Green &  'MLL-X fusions', 'TP53', 'FLT3'\\
       \hline\\
    Green light &  'Other Tyr kinases', 'MYH11.CBFB', 'DNMT3A', 'MLL-X fusions' \\
     \hline
\end{tabular}
}

%
%     Orange= ['PML.RARA', 'MYH11.CBFB', 'RUNX1.RUNX1T1', 'DNMT3A'], Orange light=['Other Tyr kinases', 'MYH11.CBFB', 'MLL.PTD', 'NPM1'], 
%    Green=['MLL-X fusions', 'TP53', 'FLT3'], 
%    Green light=['Other Tyr kinases', 'MYH11.CBFB', 'DNMT3A', 'MLL-X fusions'],
%    Blue=['PML.RARA', 'MYH11.CBFB', 'RUNX1.RUNX1T1', 'TP53', 'NPM1', 'RUNX1'],  
%    Blue light=['PML.RARA', 'MYH11.CBFB', 'RUNX1.RUNX1T1', 'TP53', 'NPM1', 'MLL.PTD'].  

\subsection{Glioblastoma multiforme data}\label{SectionGBM}
The second mutation data is taken from \cite{GBM2008}.  It has 84 patients and around 100 genes.
Approximately, 70\% of the genes have very low coverage so we removed them from the data; precisely,
we have removed all genes with fewer than 10 patients. We have used the complex
$K_\eta$, with  $\varepsilon$ fixed to 7 because lower values don't exhibit stable topologies; that is, barcodes are essentially one point long. This is a different choice of complex from the complex $K_\varepsilon$ that we used with the previous AML data, which interestingly doesn't yield noticeable stability--a possible explanation of this is the small size of the data (i.e., number of patients).   Recall that the definitions of the two complexes are given on page \pageref{complexDefs} (definitions 2 and 3). In the table below, one can see that the topology stabilizes for the first three values of $\eta$.  Any choice of pathways within this range is  considered robust (proportionate to the small size of the data). 

\begin{center}
\begin{tabular}{cccc}
	$\eta$ &   $|\mathcal N_\eta|$ &  $density(\mathcal N_\eta)$  & $\beta_i$\\
	\hline\\
				66 &							15	& 						0.73		&	 $1, 2,0, \cdots$ \\
	\hline\\
				67 &							14	& 						0.74		&	 $1, 2,0, \cdots$ \\
	\hline\\
				68 &							12	& 						0.72		&	 $1, 2,0, \cdots$ \\		
	\hline\\
				69 &							6	& 						0.6		&	 $1, 0,0, \cdots$ \\	
	\hline\\
				70 &							50	& 						1		&	 $1, 0,0, \cdots$ \\			
\end{tabular}
\end{center}
   	 
\begin{figure}[h]\label{exples}
\begin{center}
      \includegraphics[scale=0.5]{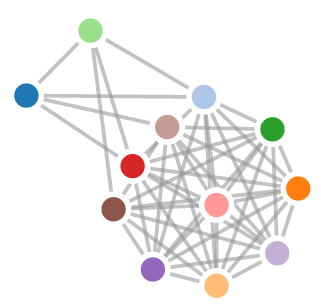} 
      \end{center}
    \caption{{\small The nerve for the space of pathways for GBM data (see table below for the legend). The barcodes are stable in the first part of the table, which indicates that the nerve  is homotopy equivalent to a genus-2 surface.}
}
\label{GBM}
\end{figure}

~~\\
The following table provides the legend for the Figure \ref{GBM} corresponding to the~{GBM data:}

~~\\
\begin{center}
{\footnotesize
\begin{tabular}{l|l}
  {\bf Color} & {\bf Genes in the pathway}\\
 \hline\\
 blue  & RB1,  NF1,  CYP27B1,  CDKN2B \\ \hline\\
 blue light & RB1,  NF1,  MDM2,  AVIL-CTDSP2,  CDKN2B\\\hline\\
 orange & TP53,  MDM2,  OS9,  CDKN2A\\\hline\\
 orange light & TP53,  MDM2,  AVIL-CTDSP2,  CDKN2A\\\hline\\
 green& TP53,  MDM2,  DTX3,  CDKN2A\\\hline\\
  green light & RB1,  NF1,  CDK4,  CDKN2B\\\hline\\
  brown& TP53,  CDK4,  CDKN2A\\\hline\\
  brown light& TP53,  CYP27B1,  CDKN2A\\\hline\\
   purple& TP53,  MDM2,  AVIL-CTDSP2,  MTAP\\\hline\\
   purple light & TP53,  MDM2,  DTX3,  MTAP\\\hline\\
  red& RB1,  NF1,  MDM2,  OS9,  CDKN2B\\\hline\\
   pink& TP53,  MDM2,  OS9,  MTAP\\\hline
 \end{tabular}
 } 
 \end{center}

\section{Conclusion}
The main goal of this paper was to suggest a study of the {\em space of cancer pathways}, using the natural language of algebraic topology. % We did not discuss here how the pathways are computed efficiently, nor how the topological objects described here are actually calculated; the formulations and algorithms are subjects of a companion paper (\cite{alghassi_quantum_2019}).
We hope that the consideration of the pathways collectively, that is, as a topological space,  helps to reveal novel relations between these pathways.  Indeed, we have seen that the homology in
the case of AML  indicates that the mutation data has the shape of a sphere.\footnote{Using a different visual representation, Vogelstein found AML to be very different from other cancers. Indeed, that has allowed many scientists to speculate that such genetically simple tumors are more susceptible to drugs, and thus intrinsically more curable.}  However, in the case of GBM, the final set of pathways  has the  topology of
a double torus (or, more technically, a genus-2 surface). 
This intriguing observation raises the question of whether these facts translate into a new biological understanding about cancer.  Studying the space of pathways of other cancers will be illuminating as well, if they also show similar structures, and we can classify cancers by the topology of their mutated driver pathways.  This is an example of the new type of hypotheses one can now  formulate about the data. Eventually, our goal (recalling Poincare) is to help build a house by revealing patterns among the stones.
Let us close with a quote from {\em The Emperor of All Maladies} (page 458):
\begin{quote}
    The third,\footnote{The first is targeted therapy on the mutated pathways,  as we mentioned in the Introduction. The second is cancer prevention through identifying preventable carcinogens.} and arguably most complex, new direction for cancer medicine is to integrate our understanding of aberrant genes and pathways to explain the {\em behavior} of cancer as a whole, thereby renewing the cycle of knowledge, discovery and therapeutic intervention.
\end{quote}

\newpage
\appendix

\section{Appendix}\label{SectionAppendix}
This appendix has three parts. In the first part, we review an efficient procedure for computing homology groups. The remaining
two parts give the obtained list of pathways for GBM  and  AML data, respectively.  

\subsection{Computing homology on quantum computers}   
We provide the following material (from \cite{raoufHomology}) for easy access.
We review  how the homology spaces $H_*(X)$ are computed.  In principle, the formal definitions of homology that can be found in any algebraic topology textbook, are sufficient for computations.  However, we point here to a more efficient approach based on Mayer-Vietoris blow-up complexes. We 
formulate finding optimal  Mayer-Vietoris blow-up complexes as  Quadratic Unconstrained Binary Optimizations (QUBO). QUBO-solvers (such as the  D-Wave quantum annealer \cite{natueDwave} or quantum-inspired classical solvers such as \cite{Hedayat2}) are now available for calculations. 
\\~~

\begin{figure}[!htbp]\label{bl1}
%\vskip 0.2in
\begin{center}
\centerline{
\includegraphics[scale=0.3]{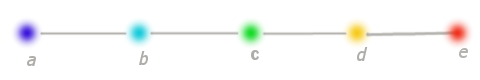}}
\centerline{
\includegraphics[scale=0.3]{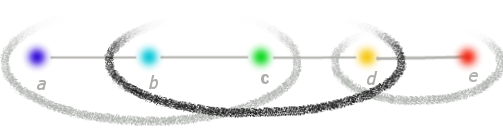}}
\centerline{
\includegraphics[scale=0.3]{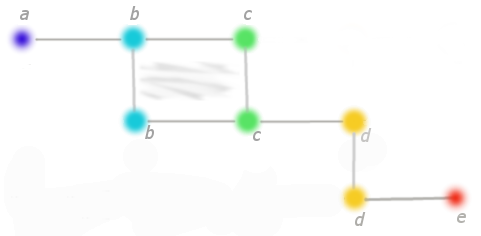}}
\caption{(Top) The simplicial complex $K$ is the depicted graph.
(Middle)  $K$ is covered with $K_0, \, K_1 $, and $K_2$. 
 %and hence coincides with its 1-skeleton. 
 % Only the zeroth Betti number is non zero and is equal to 1. 
%(Middle) The nerve can be used to brings down the top picture graph to an edge. 
(Bottom) The blow-up complex of the cover depicted in the middle picture.  After the blow-up, the edges $b\otimes 01, \, c\otimes 01, \, d\otimes 12$, and the triangle
$bc\otimes 01$ appear.}
\label{mv1}
\end{center}
%\vskip -0.2in
\end{figure}

\begin{figure}[!htbp]\label{bl2}
%\vskip 0.2in
\begin{center}
%\centerline{
%\includegraphics[scale=0.3]{pics/example2_.png}}
\centerline{
\includegraphics[scale=0.3]{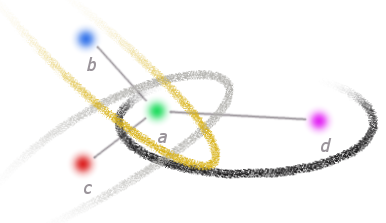}}
\centerline{
\includegraphics[scale=0.3]{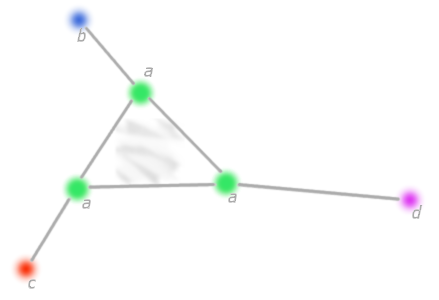}}
\caption{The triangle $a\times 012$ appears after blowing-up the cover of the middle picture.}
\label{mv2}
\end{center}
%\vskip -0.2in
\end{figure}

 Let $\mathcal C= \{K^i\}_{i\in I}$ be a cover of  $K$ by simplicial subcomplexes  $K^i\subseteq K$; here again we have dropped the subscript $\varepsilon$ from $K_\varepsilon$ for simplicity.   For $J\subseteq I,$ we define $K^J=\cap_{j\in J} K^j.$ 
\begin{definition}
The \emph{Mayer-Vietoris blow-up complex} of the simplicial complex $K$ and cover $\mathcal C,$ is defined by:
$$
	K^\mathcal C = \bigcup _{J\subseteq I} \bigcup_{ \sigma\in K^J}  \sigma\times J.
$$
A basis for the $k-$chains $C_k(K^\mathcal C)$ is {$\{ \sigma\otimes J \in K^\mathcal C |\, \mathrm{dim}\,  \sigma + \mathrm{card}\, J= n \}$}. The boundary of a cell $ \sigma\otimes J$ is given by:
$	
{	\partial ( \sigma\otimes J) = \partial  \sigma\otimes J + (-1)^{\mathrm dim\,  \sigma}  \sigma\otimes \partial J}.
$
\end{definition}
Simply put, the simplicial complex $K^\mathcal C$ is  the set of the ``original" simplices  in addition to the ones we get by blowing up common simplices.
These are of the form $\sigma\otimes J$ in the definition  above.  
In Figure~\ref{mv1}, the yellow vertex  $d$ common to the two subcomplexes $\{K_1, K_2\}$ is blown-up into an edge $d\otimes 12$,  and the edge  $bc$ is blown-up into the ``triangle"
$bc\otimes 01$. In Figure \ref{mv2}, 
the vertex $a$ common to three subcomplexes $\{K_0, K_1, K_2\}$ is blown-up into the  triangle $a\otimes 012$. 

%  A vertex appearing in two covering subcomplexes is blown-up into an edge, 
%a vertex common to three subcomplexes is blown-up into a triangle, an edge is blow-up into a triangle and so on. 

We will not prove it  here, but the projection $K^\mathcal C\rightarrow K$ is a homotopy equivalence and  induces 
an isomorphism $H_*(K^\mathcal C)\simeq H_*(K)$ \cite{Zomorodian2008126}.
 The key point is that the boundary map of the simplicial complex $K^\mathcal C$ (which replaces $K$ by the homotopy equivalence) has a nice block form suitable for parallel rank
computation. As an example, let us consider   again 
the simplicial complex $K$ depicted in Figure \ref{mv1}. First, $C_0(K^\mathcal C)$, the space  of vertices of the blow-up complex $K^\mathcal C$,  is spanned by the vertices 
$$\{a\otimes 0,\, b\otimes 0,\, c\otimes 0,\, b\otimes1, \,c\otimes 1, \,d\otimes 1, \,d\otimes 2, \,e\otimes 2\},$$
%$$C_0(K^\mathcal C) =\langle a\otimes 0,\, b\otimes 0,\, c\otimes 0,\, b\otimes1, \,c\otimes 1, \,d\otimes 1, \,d\otimes 2, \,e\otimes 2\rangle,$$
that is, all vertices of $K$ taking into account the partition they belong to.
The space of edges $C_1(K^\mathcal C)$ is spanned by 
$$\{ ab\otimes 0,  bc\otimes0, bc\otimes 1, cd\otimes 1, de\otimes 2,b\otimes 01,  \, c\otimes 01, \, d\otimes 12\},$$
%\begin{eqnarray}\nonumber
%C_1(K^\mathcal C) = &\langle ab\otimes 0,  bc\otimes0, bc\otimes 1, cd\otimes 1, de\otimes 2,b\otimes 01, \\\nonumber
%& \, c\otimes 01, \, d\otimes 12\rangle,
%\end{eqnarray}  
which is the set of the ``original" edges (edges of the form $ \sigma\otimes j$ with $j\in J=\{0, 1, 2\}$ and $ \sigma$ is an edge in $K$) and 
the new ones resulting from blow-ups, that is, those of the form $v\otimes ij$ where $v$ is a vertex
in $K^j\cap K^j$ (if the intersection is empty, the value of boundary map is just 0).
The matrix of the boundary map $\partial_0$ with respect to the given ordering is then:

 $$
 \partial_0 = 
\left(
\begin{array}{cc|cc|c|ccc}
\rowcolor{red!20}
  -1 & 0 & 0 & 0 & 0& 0&  0& 0 \\ 
  \rowcolor{red!20}
   1 & -1 & 0 & 0 & 0&-1& 0&0\\
   \rowcolor{red!20}
   0 & 1 & 0 & 0 & 0 & 0& -1&0\\
   \hline
   % \Xhline %{\arrayrulewidth}
    \rowcolor{green!20}
   0 & 0 & -1& 0 & 0& 1& 0& 0 \\ 
   \rowcolor{green!20}
  0 & 0 & 1 & -1 & 0&0& 1&0\\
  \rowcolor{green!20}
   0 & 0 & 0 & 1 & 0&0& 0&-1\\
   \hline
    %\Xhline%{\arrayrulewidth}
    \rowcolor{blue!20}
     0 & 0 & 0 & 0 & -1&0& 0&1\\
     \rowcolor{blue!20}
     0 & 0 & 0 & 0 & 1&0& 0&0\\
\end{array}
\right)
 $$
 Clearly, one can now row-reduce each coloured block independently. There might be {\it remainders}, that is, zero rows except for the intersection part. We collect all such rows in one extra matrix and row-reduce it at the end and aggregate. 
 For the second boundary matrix we need to determine $C_2(K^\mathcal C)$. The 2-simplices
 are of three forms. First, the original ones (those of the form $ \sigma\otimes j$ with $ \sigma\in C_2(K)$; in this example there is none) then those of the form $ \sigma\times \{i, j\}$, with $ \sigma$ being in $K^i\cap K^j$. And finally,  those of the form $v\otimes \{i, j, k\}$, with $v\in K^i\cap K^j\cap K^k$ (there is none in this example, but Figure 5 has one). We get 
 $
 {C_2(K^\mathcal C) =\langle bc\otimes 01\rangle}
 $; thus, there is  no need for parallel computation. 
 
 ~~\\
Finally, for efficient computations, it is necessary that the homologies of the smaller blocks are easy to compute. This is where the QUBO-solver comes in. It is used to obtain a ``good" cover $\mathcal C$   by computing  a clique cover of the original graph $G$ (plus a completion step). It is easy to see that such coverings do, indeed, come with trivial block homologies \cite{raoufHomology}.

% \subsection*{Quantum computation of $H_*(K_\varepsilon)$ } 
%To compute  the homology  of  the persistent pathway  complex  $K_\varepsilon$
%we will use the quantum algorithm \cite{raoufNerve} (we might also use \cite{raoufHomology} but  \cite{raoufNerve} is ``optimal").   In this case, 
%$$
%	H_*(G_\varepsilon)=H_*(K_\varepsilon) = H_*(\mathcal N_\varepsilon),
%$$  
%where $\mathcal N_\varepsilon$ is the nerve of an edge clique cover of $G_\varepsilon$. The passage to the edge clique cover nerve 
%gives an exponential reduction of the size of the problem.  Indeed,  the size of the 1-skeleton of $\mathcal N_\varepsilon$  is 
%of order $O(log(n))$, with $n$ being the number of genes = vertices of $G_\varepsilon$. 
%We refer the interested reader to the paper \cite{raoufNerve} for a detailed account.
%%\section*{Co-occurence}
%%{\color{blue}	 
%%- co-occurrence = define the graph $G$ for the space of pathways: two pathways are connected if they have a least a patient in common. Co-Occurrences are the cliques of this graph. We get
%%a clique complex ! and we can do this for each patient (in addition to persitence complex)
%%}

\subsection{GBM pathways list}
	 The table below reports the final pathways in Figure \ref{GBM}. 
	 Details on the data,  the parameters,  and how the pathways are computed using persistent homology, can be found in Subsection \ref{SectionGBM} on page \pageref{SectionGBM}. %Results for AML data, which are very long, are presented next 
\begin{center}	 
\begin{small}	 
\begin{tabular}{l|c}
	 {\bf Pathway} & {\bf coverage $\eta$}\\
	 \hline\\
       RB1 ,  NF1 ,  CYP27B1 ,  CDKN2B   &  70\\ 
      RB1 ,  NF1 ,  MDM2 ,  AVIL-CTDSP2 ,  CDKN2B   &  69\\ 
      TP53 ,  MDM2 ,  OS9 ,  CDKN2A   &  69\\ 
      TP53 ,  MDM2 ,  AVIL-CTDSP2 ,  CDKN2A   &  69\\ 
      TP53 ,  MDM2 ,  DTX3 ,  CDKN2A   &  69\\ 
      RB1 ,  NF1 ,  CDK4 ,  CDKN2B   &  69\\ 
      RB1 ,  NF1 ,  MDM2 ,  OS9 ,  CDKN2B   &  68\\ 
      TP53 ,  MDM2 ,  OS9 ,  MTAP   &  68\\ 
      TP53 ,  MDM2 ,  AVIL-CTDSP2 ,  MTAP   &  68\\ 
      TP53 ,  MDM2 ,  DTX3 ,  MTAP   &  68\\ 
      TP53 ,  CDK4 ,  CDKN2A   &  68\\ 
      TP53 ,  CYP27B1 ,  CDKN2A   &  68\\ 
      TP53 ,  CDK4 ,  MTAP   &  67\\ 
      TP53 ,  CYP27B1 ,  MTAP   &  67\\ 
      RB1 ,  NF1 ,  CYP27B1 ,  CDKN2A   &  66\\ 
      RB1 ,  NF1 ,  MDM2 ,  OS9 ,  CDKN2A   &  65\\ 
      RB1 ,  NF1 ,  MDM2 ,  AVIL-CTDSP2 ,  CDKN2A   &  65\\ 
      RB1 ,  NF1 ,  MDM2 ,  DTX3 ,  CDKN2B   &  65\\ 
      RB1 ,  NF1 ,  CDK4 ,  CDKN2A   &  65\\ 
      RB1 ,  NF1 ,  CYP27B1 ,  MTAP   &  65\\ 
      RB1 ,  NF1 ,  MDM2 ,  OS9 ,  MTAP   &  64\\ 
      RB1 ,  NF1 ,  MDM2 ,  AVIL-CTDSP2 ,  MTAP   &  64\\ 
      RB1 ,  NF1 ,  CDK4 ,  MTAP   &  64\\ 
      RB1 ,  NF1 ,  MDM2 ,  DTX3 ,  CDKN2A   &  62\\ 
      RB1 ,  NF1 ,  MDM2 ,  DTX3 ,  MTAP   &  61\\ 
      TP53 ,  MDM2 ,  AVIL-CTDSP2 ,  SEC61G   &  60\\ 
      RB1 ,  NF1 ,  EGFR ,  OS9   &  60\\ 
      TP53 ,  MDM2 ,  OS9 ,  SEC61G   &  59\\ 
      TP53 ,  MDM2 ,  DTX3 ,  SEC61G   &  59\\ 
      PTEN ,  IFNA21 ,  MDM2 ,  AVIL-CTDSP2   &  58\\ 
      TP53 ,  MDM2 ,  AVIL-CTDSP2 ,  IFNA21   &  58\\ 
      RB1 ,  NF1 ,  EGFR ,  DTX3   &  58\\ 
      PTEN ,  IFNA21 ,  CYP27B1   &  58\\   
      PTEN ,  IFNA21 ,  MDM2 ,  OS9   &  57\\ 
      TP53 ,  MDM2 ,  OS9 ,  IFNA21   &  57\\ 
      TP53 ,  MDM2 ,  DTX3 ,  IFNA21   &  57\\ 
      PTEN ,  IFNA21 ,  CDK4   &  57\\ 
      RB1 ,  NF1 ,  MDM2 ,  AVIL-CTDSP2 ,  SEC61G   &  55\\
      \end{tabular}
      \end{small}
      %\newpage
      
      \begin{small}
      \begin{tabular}{l|c} 
      PTEN ,  IFNA21 ,  MDM2 ,  DTX3   &  55\\ 
      TP53 ,  MDM2 ,  AVIL-CTDSP2 ,  ELAVL2   &  55\\ 
      RB1 ,  NF1 ,  MDM2 ,  OS9 ,  SEC61G   &  54\\ 
      TP53 ,  MDM2 ,  OS9 ,  ELAVL2   &  54\\ 
      TP53 ,  MDM2 ,  DTX3 ,  ELAVL2   &  54\\
      RB1 ,  NF1 ,  MDM2 ,  AVIL-CTDSP2 ,  IFNA21   &  53\\ 
      TP53 ,  CDK4 ,  IFNA21   &  53\\ 
      TP53 ,  CDK4 ,  ELAVL2   &  53\\ 
      TP53 ,  CYP27B1 ,  IFNA21   &  53\\ 
      TP53 ,  CYP27B1 ,  ELAVL2   &  53\\ 
      RB1 ,  NF1 ,  MDM2 ,  OS9 ,  IFNA21   &  52\\ 
      RB1 ,  NF1 ,  MDM2 ,  DTX3 ,  SEC61G   &  52\\ 
      RB1 ,  NF1 ,  MDM2 ,  AVIL-CTDSP2 ,  ELAVL2   &  51\\ 
      RB1 ,  NF1 ,  CYP27B1 ,  IFNA21   &  51\\ 
      RB1 ,  NF1 ,  CYP27B1 ,  ELAVL2   &  51\\ 
      RB1 ,  NF1 ,  MDM2 ,  OS9 ,  ELAVL2   &  50\\ 
      RB1 ,  NF1 ,  CDK4 ,  IFNA21   &  50\\ 
      RB1 ,  NF1 ,  CDK4 ,  ELAVL2   &  50\\ 
      RB1 ,  NF1 ,  MDM2 ,  DTX3 ,  IFNA21   &  49\\ 
      RB1 ,  NF1 ,  MDM2 ,  DTX3 ,  ELAVL2   &  47   
\end{tabular}	 
\end{small}
\end{center}

%\newpage

\subsection{AML pathways list}\label{AMLappendix}
 The following table gives the list of all 65 pathways of $\mathcal N_1$ (Figure \ref{AMLnerve})  and their coverage. 
 Details about the data and how  the pathways are obtained using persistent homology can be found in Section 3.1.  
All pathways listed below passed our robustness test.

\textwidth=6.3in
\voffset= -2cm
\hoffset= -2cm
\textheight=23.5cm
\newpage
	
\begin{center} 
\begin{small}
\begin{tabular}{l|c}
	{\bf Pathway} & {\bf Coverage}\\
	\hline\\
    PML.RARA ,  MYH11.CBFB ,  RUNX1.RUNX1T1 ,  TP53 ,  NPM1 ,  RUNX1   & 
  123\\  
    PML.RARA ,  MYH11.CBFB ,  RUNX1.RUNX1T1 ,  TP53 ,  NPM1 ,  MLL.PTD   & 
  113\\  
    PML.RARA ,  MYH11.CBFB ,  RUNX1.RUNX1T1 ,  DNMT3A  &  85\\  
    Other Tyr kinases ,  MYH11.CBFB ,  MLL.PTD ,  NPM1  &  83\\  
    MLL-X fusions ,  TP53 ,  FLT3  &  81\\  
    Other Tyr kinases ,  MYH11.CBFB ,  DNMT3A ,  MLL-X fusions   & 
  80\\  
    PML.RARA ,  MYH11.CBFB ,  Cohesin ,  Other modifiers  &  78\\  
    MLL-X fusions ,  DNMT3A ,  RUNX1.RUNX1T1 ,  MYH11.CBFB  &  78\\  
    PML.RARA ,  MYH11.CBFB ,  RUNX1.RUNX1T1 ,  TP53 ,  PHF6 ,  IDH1   & 
  75\\  
    Other myeloid TFs ,  MYH11.CBFB ,  Cohesin ,  Other modifiers   & 
  74\\  
    Other Tyr kinases ,  FLT3 ,  MLL-X fusions  &  74\\  
    PML.RARA ,  MYH11.CBFB ,  RUNX1.RUNX1T1 ,  TP53 ,  IDH2  &  70\\  
    PML.RARA ,  MYH11.CBFB ,  RUNX1.RUNX1T1 ,  CEBPA ,  RUNX1   & 
  66\\  
    PML.RARA ,  MYH11.CBFB ,  RUNX1.RUNX1T1 ,  TP53 ,  PHF6 ,  MLL.PTD   & 
  65\\  
    Other myeloid TFs ,  MYH11.CBFB ,  TP53 ,  RUNX1.RUNX1T1 ,  RUNX1   & 
  65\\  
    PML.RARA ,  PTPs ,  IDH2 ,  TET2  &  65\\  
    PML.RARA ,  MYH11.CBFB ,  TET2 ,  IDH2    &  64\\  
    MLL-X fusions ,  TP53 ,  RUNX1.RUNX1T1 ,  MYH11.CBFB ,  IDH2   & 
  63\\  
    PML.RARA ,  MYH11.CBFB ,  RUNX1.RUNX1T1 ,  CEBPA ,  PHF6 ,  MLL.PTD   & 
  62\\  
    PML.RARA ,  KRAS/NRAS ,  PHF6 ,  MLL.PTD ,  KIT  &  62\\  
    MLL-X fusions ,  TP53 ,  RUNX1.RUNX1T1 ,  MYH11.CBFB ,  IDH1   & 
  62\\  
    MLL-X fusions ,  TP53 ,  RUNX1.RUNX1T1 ,  MYH11.CBFB ,  RUNX1   & 
  62\\  
    Other myeloid TFs ,
    MYH11.CBFB ,
    TP53 ,
    RUNX1.RUNX1T1 ,
    PHF6 ,
    MLL.PTD   & 
  61\\  
    PML.RARA ,  KRAS/NRAS ,  PHF6 ,  MLL.PTD ,  RUNX1.RUNX1T1   & 
  61\\  
    PML.RARA ,  KIT ,  TP53 ,  IDH2  &  60\\  
    PML.RARA ,  KIT ,  TP53 ,  RUNX1  &  59\\  
    MLL-X fusions ,  TET2 ,  MYH11.CBFB ,  IDH2  &  57\\  
    PML.RARA ,  PTPs ,  IDH2 ,  KIT  &  56\\  
    PML.RARA ,  KIT ,  CEBPA ,  RUNX1  &  56\\  
    PML.RARA ,  KIT ,  TP53 ,  PHF6 ,  MLL.PTD  &  55\\  
    PML.RARA ,  PTPs ,  IDH2 ,  RUNX1.RUNX1T1  &  55\\  
    PML.RARA ,  PTPs ,  RUNX1 ,  KIT  &  55\\  
    Other myeloid TFs ,  KIT ,  TP53 ,  RUNX1  &  55\\  
    PML.RARA ,  PTPs ,  RUNX1 ,  RUNX1.RUNX1T1  &  54\\  
    MLL-X fusions ,  TP53 ,  KIT ,  IDH2  &  53\\  
    PML.RARA ,  KIT ,  CEBPA ,  PHF6 ,  MLL.PTD  &  52\\  
    MLL-X fusions ,  TP53 ,  RUNX1.RUNX1T1 ,  MYH11.CBFB ,  MLL.PTD   & 
  52\\  
    Ser-Tyr kinases ,  RUNX1.RUNX1T1 ,  PTPs ,  MLL.PTD  &  52\\  
    MLL-X fusions ,  TP53 ,  KIT ,  RUNX1  &  52\\  
    \end{tabular}
    \end{small}
  
    \begin{small}
    \begin{tabular}{l|c}
      	{\bf Pathway} & {\bf Coverage}\\
	\hline\\
    Other myeloid TFs ,  KIT ,  TP53 ,  PHF6 ,  MLL.PTD  &  51\\  
    PML.RARA ,  WT1 ,  RUNX1.RUNX1T1 ,  TP53  &  51\\  
    PML.RARA ,  MYH11.CBFB ,  TET2 ,  PHF6  &  50\\  
    PML.RARA ,  KIT ,  Cohesin  &  50\\
    Other Tyr kinases ,  MYH11.CBFB ,  IDH2 ,  MLL-X fusions  &  49\\  
    Other Tyr kinases ,  MYH11.CBFB ,  IDH1 ,  MLL-X fusions  &  48\\  
    Other Tyr kinases ,  MYH11.CBFB ,  MLL.PTD ,  PHF6 ,  Other myeloid TFs   & 
  47\\  
    Other Tyr kinases ,  KRAS/NRAS ,  PHF6 ,  MLL.PTD  &  47\\  
    Other myeloid TFs ,  MYH11.CBFB ,  TET2 ,  PHF6  &  46\\  
    MLL-X fusions ,  Cohesin ,  MYH11.CBFB  &  46\\  
    Other myeloid TFs ,  KIT ,  Cohesin  &  46\\  
    Other Tyr kinases ,  MYH11.CBFB ,  IDH1 ,  PHF6  &  45\\  
    PML.RARA ,  PTPs ,  MLL.PTD ,  KIT  &  45\\  
    PML.RARA ,  WT1 ,  TET2  &  45\\  
    PML.RARA ,  PTPs ,  MLL.PTD ,  RUNX1.RUNX1T1  &  44\\  
    MLL-X fusions ,  TP53 ,  RUNX1.RUNX1T1 ,  WT1  &  44\\  
    MLL-X fusions ,  Cohesin ,  KIT  &  43\\  
    MLL-X fusions ,  TP53 ,  KIT ,  MLL.PTD  &  42\\  
    Other Tyr kinases ,  PTPs ,  IDH2  &  41\\  
    Other Tyr kinases ,  MYH11.CBFB ,  MLL.PTD ,  MLL-X fusions   & 
  38\\  
    MLL-X fusions ,  TET2 ,  WT1  &  38\\  
    Spliceosome ,  CEBPA  &  37\\  
    Spliceosome ,  PTPs  &  36\\  
    Spliceosome ,  Other myeloid TFs  &  36\\  
    Other Tyr kinases ,  WT1 ,  MLL-X fusions  &  30\\  
    Other Tyr kinases ,  PTPs ,  MLL.PTD  &  30\\    
\end{tabular}
\end{small}
\end{center}

~~\\
{The following long table gives the assignment {\it tumor $\mapsto$ list of pathways} for the AML data.}
For each tumor we assign a robust (in our topological sense) set of pathways. More specifically, for a tumor (sample ID) i, the second column  gives
the list of all pathways  that have passed our robustness test. %\afterpage 

~~\\
%\afterpage
\scalebox{.9}{
   % \clearpage% Flush earlier floats (otherwise order might not be correct)
    %\thispagestyle{empty}% empty page style (?)
   % \begin{landscape}% Landscape page
        \centering % Center table
\begin{tiny}
\begin{tabular}{l|l|l}
{\bf Sample ID} & {\bf Pathways} & {\bf Cyto}\\
\hline
0 &   0, 1, 2, 6, 8, 11, 12, 13, 15, 16, 18, 19, 23, 24, 25, 27, 28, 29, 30, 31, 33, 35, 40, 41, 42, 51, 52, 53    & Good\\
1 &   0, 1, 2, 7, 8, 11, 12, 13, 14, 17, 18, 20, 21, 22, 23, 30, 33, 36, 37, 40, 53, 54    & Good\\
2 &   0, 1, 2, 3, 4, 5, 6, 7, 8, 9, 10, 11, 12, 13, 14, 16, 17, 18, 19, 20, 21, 22, 23, 26, 36, 41, 43, 44, 45, 46, 47, 48, 50, 58    & Good\\
3 &   0, 1, 2, 3, 5, 6, 7, 8, 9, 11, 12, 13, 14, 16, 17, 18, 20, 21, 22, 26, 36, 41, 43, 44, 45, 47, 48, 50, 58    & Good\\
4 &   0, 1, 2, 3, 5, 6, 8, 10, 11, 12, 13, 15, 16, 18, 19, 23, 24, 25, 27, 28, 29, 30, 31, 33, 35, 40, 41, 42, 43, 44, 45, 46, 50, 51, 52, 53, 57, 58, 63, 64    & Good\\
5 &   0, 1, 2, 4, 6, 8, 10, 11, 12, 13, 15, 16, 18, 19, 23, 24, 25, 27, 28, 29, 30, 31, 33, 35, 40, 41, 42, 51, 52, 53    & Good\\
6 &   0, 1, 2, 3, 5, 6, 7, 8, 9, 11, 12, 13, 14, 15, 16, 17, 18, 20, 21, 22, 26, 27, 30, 31, 33, 36, 37, 41, 43, 44, 45, 47, 48, 50, 51, 53, 57, 58, 61, 64    & Good\\
7 &   0, 1, 2, 3, 5, 6, 7, 8, 9, 10, 11, 12, 13, 14, 17, 18, 19, 20, 21, 22, 23, 24, 25, 27, 28, 29, 30, 31, 32, 33, 34, 35, 36, 37, 38, 39, 40, 42, 43, 44, 45, \\
& 46, 49, 50, 51, 53, 54, 55, 56, 57, 58, 60, 61, 62, 63, 64    & Good\\
8 &   0, 1, 2, 4, 7, 8, 10, 11, 12, 13, 14, 17, 18, 20, 21, 22, 23, 30, 33, 36, 37, 40, 53, 54    & Good\\
9 &   0, 1, 2, 6, 8, 11, 12, 13, 15, 16, 18, 19, 23, 24, 25, 27, 28, 29, 30, 31, 33, 35, 40, 41, 42, 51, 52, 53    & Good\\
10 &   0, 1, 2, 3, 5, 6, 7, 8, 9, 11, 12, 13, 14, 16, 17, 18, 20, 21, 22, 26, 36, 41, 43, 44, 45, 47, 48, 50, 58    & Good\\
11 &   0, 1, 2, 6, 8, 11, 12, 13, 15, 16, 18, 19, 23, 24, 25, 27, 28, 29, 30, 31, 33, 35, 40, 41, 42, 51, 52, 53, 60, 61, 62    & Good\\
12 &   40, 52, 54, 59, 60, 61, 62, 63    & Good\\
13 &   0, 1, 2, 3, 5, 6, 7, 8, 9, 11, 12, 13, 14, 16, 17, 18, 19, 20, 21, 22, 24, 25, 26, 27, 28, 29, 31, 32, 34, 35, 36, 38, 39, 41, 42, 43, 44, 45, 47, 48, 49, \\
& 50, 51, 55, 56, 58    & Good\\
14 &   0, 1, 2, 4, 6, 8, 10, 11, 12, 13, 15, 16, 18, 19, 23, 24, 25, 27, 28, 29, 30, 31, 33, 35, 40, 41, 42, 51, 52, 53    & Good\\
15 &   0, 1, 2, 6, 8, 11, 12, 13, 15, 16, 18, 19, 23, 24, 25, 27, 28, 29, 30, 31, 33, 35, 40, 41, 42, 51, 52, 53    & Good\\
16 &   0, 1, 2, 3, 5, 6, 7, 8, 9, 10, 11, 12, 13, 14, 17, 18, 20, 21, 22, 23, 30, 33, 36, 37, 40, 42, 43, 44, 45, 46, 48, 49, 50, 53, 54, 55, 57, 58, 63, 64    & Good\\
17 &   0, 1, 2, 4, 6, 8, 10, 11, 12, 13, 15, 16, 18, 19, 23, 24, 25, 27, 28, 29, 30, 31, 33, 35, 40, 41, 42, 51, 52, 53    & Good\\
18 &   0, 1, 2, 3, 5, 6, 7, 8, 9, 11, 12, 13, 14, 16, 17, 18, 19, 20, 21, 22, 24, 25, 26, 27, 28, 29, 31, 32, 34, 35, 36, 37, 38, 39, 41, 42, 43, 44, 45, 47, 48, \\
& 49, 50, 51, 55, 56, \\
& 58, 60, 61, 62    & Good\\
19 &   0, 1, 2, 6, 8, 11, 12, 13, 15, 16, 18, 19, 23, 24, 25, 27, 28, 29, 30, 31, 33, 35, 40, 41, 42, 51, 52, 53    & Good\\
20 &   0, 1, 2, 3, 5, 6, 7, 8, 9, 11, 12, 13, 14, 16, 17, 18, 20, 21, 22, 26, 36, 41, 43, 44, 45, 47, 48, 50, 58    & Good\\
21 &   0, 1, 2, 7, 8, 11, 12, 13, 14, 15, 16, 17, 18, 19, 20, 21, 22, 23, 24, 25, 26, 27, 28, 29, 30, 31, 32, 33, 34, 35, 36, 37, 38, 39, 40, 41, 42, 47, 49,\\
&  51, 52, 53, 54, 55, 56,  59, 60, 61, 62    & Good\\
22 &   32, 4, 39, 9, 10, 45, 14, 47, 49, 22, 62    & Good\\
23 &   0, 1, 2, 6, 8, 11, 12, 13, 15, 16, 18, 19, 23, 24, 25, 27, 28, 29, 30, 31, 33, 35, 40, 41, 42, 51, 52, 53    & Good\\
24 &   0, 1, 2, 3, 4, 5, 6, 7, 8, 9, 10, 11, 12, 13, 14, 16, 17, 18, 20, 21, 22, 26, 36, 41, 43, 44, 45, 47, 48, 50, 58    & Good\\
25 &   0, 1, 2, 3, 5, 6, 7, 8, 9, 11, 12, 13, 14, 16, 17, 18, 19, 20, 21, 22, 24, 25, 26, 27, 28, 29, 31, 32, 34, 35, 36, 38, 39, 40, 41, 42, 43, 44, 45, 47, 48, \\
& 49, 50, 51, 52, 54, 55, 56,  58, 59, 63    & Good\\
26 &   0, 1, 2, 3, 5, 6, 7, 8, 9, 11, 12, 13, 14, 16, 17, 18, 19, 20, 21, 22, 23, 26, 36, 41, 43, 44, 45, 46, 47, 48, 50, 58    & Good\\
27 &   0, 1, 2, 6, 8, 9, 11, 12, 13, 14, 15, 16, 18, 19, 22, 23, 24, 25, 27, 28, 29, 30, 31, 32, 33, 35, 39, 40, 41, 42, 45, 47, 49, 51, 52, 53, 62    & Good\\
28 &   0, 1, 2, 4, 6, 8, 10, 11, 12, 13, 15, 16, 18, 19, 23, 24, 25, 27, 28, 29, 30, 31, 33, 35, 40, 41, 42, 51, 52, 53    & Good\\
29 &   0, 1, 2, 6, 7, 8, 9, 11, 12, 13, 14, 17, 18, 20, 21, 22, 23, 30, 33, 36, 37, 40, 42, 48, 49, 53, 54, 55    & Good\\
30 &   0, 1, 2, 4, 5, 6, 7, 8, 10, 11, 12, 13, 15, 16, 17, 18, 19, 20, 21, 23, 24, 25, 26, 27, 28, 29, 30, 31, 33, 34, 35, 36, 38, 40, 41, 42, 43, 44, 48,\\
& 51, 52, 53, 54, 55, 56, 58, 59, 63    & Good\\
31 &   37    & Good\\
32 &   0, 1, 2, 3, 5, 6, 7, 8, 9, 11, 12, 13, 14, 16, 17, 18, 20, 21, 22, 26, 36, 41, 43, 44, 45, 47, 48, 50, 58    & Good\\
33 &   0, 1, 2, 3, 5, 6, 8, 10, 11, 12, 13, 15, 16, 18, 19, 23, 24, 25, 27, 28, 29, 30, 31, 33, 35, 40, 41, 42, 43, 44, 45, 46, 50, 51, 52, 53, 57, 58, 63, 64    & Good\\
34 &   0, 1, 2, 6, 7, 8, 9, 11, 12, 13, 14, 17, 18, 20, 21, 22, 23, 30, 33, 36, 37, 40, 42, 48, 49, 53, 54, 55    & Good\\
35 &   0, 1, 2, 4, 6, 8, 10, 11, 12, 13, 15, 16, 18, 19, 23, 24, 25, 27, 28, 29, 30, 31, 33, 35, 40, 41, 42, 51, 52, 53    & Good\\
36  &   10, 4    & Good\\
37 &   0, 1, 2, 3, 5, 7, 8, 64, 44, 61, 15, 50, 51, 20, 53, 57, 33, 27, 37, 30, 31    & Intermediate\\
38 &   35, 6, 39, 8, 41, 13, 46, 45, 50, 18, 19, 9, 22, 23, 47, 29    & Intermediate\\
39 &   0, 32, 34, 43, 38, 33, 11, 12, 14, 15, 16, 17, 21, 25, 24, 57, 26, 27, 28, 30, 31    & Intermediate\\
40 &   0, 11, 12, 14, 15, 16, 17, 21, 24, 25, 26, 27, 28, 30, 31, 32, 33, 34, 37, 38, 43, 57    & Intermediate\\
41 &   35, 60, 12, 46, 18, 19, 23, 28    & Intermediate\\
42 &   0, 1, 2, 3, 5, 7    & Intermediate\\
43 &   0, 1, 2, 3, 4, 5, 6, 7, 9, 10, 48, 49, 55, 42    & Intermediate\\
44 &   0, 6, 9, 12, 14, 15, 16, 19, 21, 23, 25, 26, 28, 31, 32, 33, 38, 41, 42, 46, 47, 48, 49, 52, 55, 59, 60, 61, 62    & Intermediate\\
45 &   0, 1, 10, 3, 4    & Intermediate\\
46 &   2, 43, 5, 7, 11, 34, 46, 15, 16, 17, 19, 23, 24, 57, 26, 27, 30, 37    & Intermediate\\
47 &   10, 4    & Intermediate\\
48 &   0, 1, 3, 4, 6, 9, 10, 15, 27, 30, 31, 33, 37, 42, 48, 49, 51, 53, 55, 57, 61, 64    & Intermediate\\
49 &   0, 1, 2, 3, 4, 5, 7, 10    & Intermediate\\
50 &   0, 1, 2, 3, 4, 5, 7, 41, 10, 15, 16, 59, 52, 26, 47    & Intermediate\\
51 &   0, 1, 3, 4, 6, 9, 10, 15, 27, 30, 31, 33, 37, 42, 48, 49, 51, 53, 55, 57, 61, 64    & Intermediate\\
52 &   0, 1, 3, 37, 6, 33, 64, 9, 42, 15, 48, 49, 51, 53, 55, 57, 27, 61, 30, 31    & Intermediate\\
53 &   2, 59, 5, 7, 41, 15, 16, 52, 26, 47    & Intermediate\\
54 &   0, 1, 2, 3, 4, 5, 7, 10, 46, 19, 23    & Intermediate\\
55 &   0, 1, 2, 3, 4, 5, 6, 7, 9, 10, 48, 49, 55, 42    & Intermediate\\
56 &   0, 1, 2, 3, 4, 5, 7, 10    & Intermediate\\
57 &   1, 3, 8, 13, 18, 19, 20, 22, 23, 29, 35, 36, 37, 39, 44, 45, 46, 50, 51, 53, 56, 58, 64    & Intermediate\\
58 &   0, 2, 5, 6, 7, 8, 9, 12, 13, 14, 18, 19, 21, 22, 23, 25, 28, 29, 31, 32, 33, 35, 37, 38, 39, 41, 42, 45, 46, 47, 48, 49, 50, 55, 60, 61, 62    & Intermediate\\
59 &   2, 5, 7, 46, 19, 23, 60, 61, 62    & Intermediate\\
60 &   0, 1, 3, 4, 6, 9, 10, 48, 49, 55, 42    & Intermediate\\
61 &   0, 1, 3, 4, 6, 40, 9, 10, 48, 49, 52, 54, 55, 59, 42, 63    & Intermediate\\
62 &   10, 4    & Intermediate\\
63 &   28, 18, 35, 12, 60    & Intermediate\\
64 &   0, 8, 11, 12, 14, 15, 16, 17, 20, 21, 24, 25, 26, 27, 28, 30, 31, 32, 33, 34, 38, 43, 44, 50, 57, 60, 61, 62    & Intermediate\\
65 &   48, 2, 59, 5, 6, 7, 8, 41, 42, 55, 44, 15, 16, 49, 50, 52, 9, 20, 26, 47    & Intermediate\\
66 &   32, 2, 5, 7, 9, 39, 45, 14, 47, 49, 22, 62    & Intermediate\\
67 &   0, 1, 2, 3, 5, 6, 7, 9, 42, 48, 49, 55    & Intermediate\\
68 &   0, 1, 2, 3, 4, 5, 7, 10    & Intermediate\\
69 &   0, 1, 2, 3, 5, 6, 7, 40, 9, 44, 8, 50, 52, 54, 20, 59, 63    & Intermediate\\
70 &   0, 1, 34, 3, 11, 46, 15, 16, 17, 19, 43, 23, 24, 57, 26, 27, 30    & Intermediate\\
71 &   32, 39, 9, 45, 14, 47, 49, 22, 62    & Intermediate\\
72 &   0, 1, 34, 3, 4, 10, 11, 15, 16, 17, 43, 24, 57, 26, 27, 30    & Intermediate\\
73 &   0, 1, 3, 8, 44, 50, 20    & Intermediate\\
74 &   0, 1, 3, 4, 41, 10, 15, 16, 59, 52, 26, 47    & Intermediate\\
75 &   35, 4, 60, 10, 12, 18, 28    & Intermediate\\
76 &   0, 1, 2, 3, 4, 5, 7, 10    & Intermediate\\
77 &   1, 2, 3, 4, 5, 6, 7, 9, 10, 13, 15, 16, 18, 19, 22, 23, 26, 29, 35, 36, 37, 39, 41, 45, 46, 47, 51, 52, 53, 56, 58, 59, 64    & Intermediate\\
78 &   10, 4    & Intermediate\\
79 &   2, 5, 7    & Intermediate\\
80 &   0, 1, 2, 3, 4, 5, 7, 9, 10, 39, 45, 14, 47, 49, 32, 22, 62    & Intermediate\\
81 &   34, 4, 5, 38, 7, 10, 43, 44, 48, 17, 20, 21, 54, 55, 56, 36, 26, 59, 58, 63    & Intermediate\\
82 &   4, 6, 8, 9, 10, 44, 50, 20    & Intermediate\\
83 &   2, 5, 7, 9, 14, 15, 22, 27, 30, 31, 32, 33, 37, 39, 45, 47, 49, 51, 53, 57, 61, 62, 64    & Intermediate\\
\end{tabular}
\end{tiny}
    %\end{landscape}
    %\clearpage% Flush page
}%

%\newpage
{%
    %\clearpage% Flush earlier floats (otherwise order might not be correct)
    %\thispagestyle{empty}% empty page style (?)
   % \begin{landscape}% Landscape page
        \centering % Center table
        \begin{tiny}
\begin{tabular}{c|l|l}
84 &   0, 1, 3, 4, 10, 12, 13, 14, 18, 19, 21, 22, 23, 25, 28, 29, 31, 32, 33, 35, 36, 37, 38, 39, 40, 45, 46, 51, 52, 53, 54, 56, 58, 59, 63, 64    & Intermediate\\
85 &   9, 10, 4, 6    & Intermediate\\
86 &   0, 1, 3    & Intermediate\\
87 &   0, 1, 10, 3, 4    & Intermediate\\
88 &   34, 43, 37, 11, 15, 16, 17, 62, 24, 57, 26, 27, 60, 61, 30    & Intermediate\\
89 &   0, 1, 3    & Intermediate\\
90 &   0, 1, 2, 3, 5, 7, 8, 64, 44, 61, 15, 50, 51, 20, 53, 57, 33, 27, 37, 30, 31    & Intermediate\\
91 &   0, 1, 2, 3, 5, 6, 7, 9, 12, 15, 18, 19, 23, 27, 28, 30, 31, 33, 35, 37, 40, 42, 46, 48, 49, 51, 52, 53, 54, 55, 57, 59, 60, 61, 63, 64    & Intermediate\\
92 &   0, 32, 34, 43, 38, 33, 11, 12, 14, 15, 16, 17, 21, 25, 24, 57, 26, 27, 28, 30, 31    & Intermediate\\
93 &   59, 37, 41, 47, 16, 52, 26, 15, 60, 61, 62    & Intermediate\\
94 &   0, 1, 2, 3, 4, 5, 7, 8, 9, 10, 39, 44, 45, 14, 47, 49, 50, 20, 32, 22, 62    & Intermediate\\
95 &   0, 1, 2, 3, 5, 6, 7, 9, 42, 55, 46, 48, 49, 19, 23    & Intermediate\\
96 &   0, 32, 38, 33, 12, 14, 21, 25, 28, 31    & Intermediate\\
97 &   59, 41, 46, 47, 16, 19, 52, 23, 26, 15    & Intermediate\\
98 &   0, 1, 10, 3, 4    & Intermediate\\
99 &   0, 1, 3, 37, 8, 44, 50, 20    & Intermediate\\
100 &   0, 1, 2, 4, 5, 7, 8, 11, 13, 14, 17, 20, 21, 22, 24, 25, 29, 32, 34, 36, 37, 38, 39, 40, 54, 56    & Intermediate\\
101 &   28, 18, 35, 12, 60    & Intermediate\\
102 &   0, 1, 3, 8, 44, 50, 20    & Intermediate\\
103 &   0, 32, 38, 33, 12, 14, 21, 25, 28, 31    & Intermediate\\
104 &   34, 4, 5, 38, 7, 10, 43, 44, 48, 17, 20, 21, 54, 55, 56, 36, 26, 59, 58, 63    & Intermediate\\
105 &   32, 51, 34, 35, 38, 39, 60, 42, 61, 49, 19, 56, 55, 24, 25, 27, 28, 29, 62, 31    & Intermediate\\
106 &   0, 1, 34, 3, 37, 11, 15, 16, 17, 43, 24, 57, 26, 27, 30    & Intermediate\\
107 &   0, 3, 5, 10, 12, 14, 21, 25, 28, 31, 32, 33, 38, 43, 44, 45, 46, 50, 57, 58, 63, 64    & Intermediate\\
108 &   4, 5, 6, 7, 9, 10, 12, 17, 18, 20, 21, 26, 28, 34, 35, 36, 38, 43, 44, 48, 54, 55, 56, 58, 59, 60, 63    & Intermediate\\
109 &   3, 5, 10, 12, 15, 16, 18, 26, 28, 35, 41, 43, 44, 45, 46, 47, 50, 52, 57, 58, 59, 60, 63, 64    & Intermediate\\
110 &   4, 40, 10, 52, 54, 59, 63    & Intermediate\\
111 &   0, 1, 2, 3, 4, 5, 7, 10    & Intermediate\\
112 &   0, 1, 3, 4, 5, 6, 7, 9, 10, 17, 19, 20, 21, 23, 26, 34, 36, 38, 43, 44, 46, 48, 54, 55, 56, 58, 59, 63    & Intermediate\\
113 &   0, 1, 3    & Intermediate\\
114 &   0, 1, 3, 12, 13, 14, 18, 19, 21, 22, 23, 25, 28, 29, 31, 32, 33, 35, 36, 37, 38, 39, 45, 46, 51, 53, 56, 58, 60, 61, 62, 64    & Intermediate\\
115 &   0, 1, 2, 3, 5, 7, 46, 19, 23    & Intermediate\\
116 &   1, 2, 3, 4, 5, 7, 10, 11, 13, 15, 16, 17, 18, 19, 22, 23, 24, 26, 27, 29, 30, 34, 35, 36, 37, 39, 43, 45, 46, 51, 53, 56, 57, 58, 64    & Intermediate\\
117 &   2, 4, 37, 6, 7, 9, 10, 61, 48, 49, 55, 60, 42, 62, 5    & Intermediate\\
118 &   0, 1, 2, 3, 5, 7    & Intermediate\\
119 &   0, 1, 3, 40, 52, 54, 59, 60, 61, 62, 63    & Intermediate\\
120 &   0, 32, 34, 43, 38, 33, 11, 12, 14, 15, 16, 17, 21, 25, 24, 57, 26, 27, 28, 30, 31    & Intermediate\\
121 &   34, 4, 37, 38, 7, 63, 10, 43, 44, 48, 17, 20, 21, 54, 55, 56, 36, 26, 59, 58, 5    & Intermediate\\
122 &   0, 1, 3, 4, 8, 10, 13, 18, 19, 22, 23, 29, 35, 39, 40, 41, 45, 46, 47, 50, 52, 54, 59, 63    & Intermediate\\
123 &   0, 1, 2, 3, 4, 5, 6, 7, 9, 10, 48, 49, 55, 42    & Intermediate\\
124 &   0, 1, 3, 8, 44, 46, 50, 19, 20, 23    & Intermediate\\
125 &   48, 59, 4, 6, 41, 10, 55, 47, 16, 49, 52, 9, 26, 15, 42    & Intermediate\\
126 &   0, 1, 2, 3, 4, 5, 7, 10    & Intermediate\\
127 &   1, 2, 3, 5, 6, 7, 9, 13, 18, 19, 22, 23, 29, 35, 36, 37, 39, 42, 45, 46, 48, 49, 51, 53, 55, 56, 58, 64    & Intermediate\\
128 &   0, 32, 38, 33, 12, 14, 21, 25, 28, 31    & Intermediate\\
129 &   1, 3, 6, 9, 13, 15, 16, 18, 19, 22, 23, 26, 29, 35, 36, 37, 39, 41, 42, 45, 46, 47, 48, 49, 51, 52, 53, 55, 56, 58, 59, 64    & Intermediate\\
130 &   28, 18, 35, 12, 60    & Intermediate\\
131 &   2, 43, 5, 6, 7, 9, 11, 34, 15, 16, 17, 24, 57, 26, 27, 30    & Intermediate\\
133 &   0, 1, 2, 3, 4, 5, 7, 10    & Intermediate\\
134 &   0, 6, 8, 9, 12, 13, 14, 18, 19, 21, 22, 23, 25, 28, 29, 31, 32, 33, 35, 38, 39, 40, 41, 45, 46, 47, 50, 52, 54, 59, 63    & Intermediate\\
135 &   32, 2, 35, 5, 7, 60, 39, 12, 45, 14, 47, 49, 18, 22, 9, 28, 62    & Intermediate\\
136 &   3, 5, 6, 9, 10, 15, 16, 26, 37, 41, 43, 44, 45, 46, 47, 50, 52, 57, 58, 59, 63, 64    & Intermediate\\
137 &   4, 5, 7, 9, 10, 14, 17, 19, 20, 21, 22, 23, 26, 32, 34, 36, 38, 39, 43, 44, 45, 46, 47, 48, 49, 54, 55, 56, 58, 59, 62, 63    & Intermediate\\
138 &   1, 3, 11, 13, 15, 16, 17, 18, 19, 22, 23, 24, 26, 27, 29, 30, 34, 35, 36, 37, 39, 43, 45, 46, 51, 53, 56, 57, 58, 64    & Intermediate\\
140 &   0, 1, 2, 3, 4, 5, 6, 7, 9, 10, 15, 27, 30, 31, 33, 37, 42, 48, 49, 51, 53, 55, 57, 61, 64    & Intermediate\\
141 &   0, 1, 2, 3, 5, 6, 7, 9, 60, 61, 62    & Intermediate\\
142 &   0, 32, 38, 33, 12, 14, 21, 25, 28, 31    & Intermediate\\
143 &   0, 1, 2, 6, 8, 11, 12, 13, 15, 16, 18, 19, 23, 24, 25, 27, 28, 29, 30, 31, 33, 35, 37, 40, 41, 42, 51, 52, 53    & Intermediate\\
144 &   0, 6, 8, 9, 11, 12, 13, 14, 15, 16, 17, 18, 19, 21, 22, 23, 24, 25, 26, 27, 28, 29, 30, 31, 32, 33, 34, 35, 38, 39, 41, 43, 45, 46, 47, 50, 57    & Intermediate\\
145 &   2, 43, 5, 7, 11, 34, 46, 15, 16, 17, 19, 23, 24, 57, 26, 27, 30    & Intermediate\\
146 &   0, 1, 3, 4, 6, 60, 10, 55, 12, 48, 49, 18, 35, 9, 28, 42    & Intermediate\\
147 &   0, 32, 59, 37, 38, 33, 31, 41, 12, 46, 47, 16, 19, 52, 14, 23, 25, 26, 15, 28, 21    & Intermediate\\
148 &   0, 1, 3    & Intermediate\\
149 &   37    & Intermediate\\
150 &   1, 3, 11, 13, 15, 16, 17, 18, 19, 22, 23, 24, 26, 27, 29, 30, 34, 35, 36, 37, 39, 43, 45, 46, 51, 53, 56, 57, 58, 64    & Intermediate\\
151  &   2, 5, 7, 8, 44, 50, 20    & Intermediate\\
152 &   0, 1, 34, 3, 11, 15, 16, 17, 43, 24, 57, 26, 27, 30    & N.D.\\
153 &   0, 1, 3, 4, 5, 8, 10, 11, 13, 14, 17, 20, 21, 22, 24, 25, 29, 32, 34, 36, 38, 39, 40, 43, 44, 45, 46, 50, 54, 56, 57, 58, 63, 64    & N.D.\\
154 &   0, 1, 2, 3, 4, 5, 6, 7, 9, 10, 60, 61, 62    & N.D.\\
155 &   0, 6, 9, 11, 12, 14, 15, 16, 17, 21, 24, 25, 26, 27, 28, 30, 31, 32, 33, 34, 38, 42, 43, 48, 49, 55, 57    & N.D.\\
157 &   0, 1, 2, 4, 5, 6, 7, 8, 9, 11, 13, 14, 15, 16, 17, 20, 21, 22, 24, 25, 26, 29, 32, 34, 36, 37, 38, 39, 40, 41, 42, 47, 48, 49, 52, \\ & 54, 55, 56, 59    & Poor\\
158 &   9, 10, 4, 6    & Poor\\
159 &   0, 1, 4, 8, 11, 13, 14, 17, 20, 21, 22, 24, 25, 29, 32, 34, 36, 38, 39, 40, 54, 56, 60, 61, 62    & Poor\\
160 &   0, 1, 4, 8, 11, 13, 14, 17, 20, 21, 22, 24, 25, 29, 32, 34, 36, 38, 39, 40, 54, 56    & Poor\\
161 &   10, 4    & Poor\\
162 &   0, 1, 3, 4, 37, 10    & Poor\\
163 &   32, 39, 9, 45, 14, 47, 49, 22, 62    & Poor\\
164 &   9, 11, 12, 14, 15, 16, 17, 18, 22, 24, 26, 27, 28, 30, 32, 34, 35, 37, 39, 40, 43, 45, 47, 49, 52, 54, 57, 59, 60, 62, 63    & Poor\\
165 &   37    & Poor\\
166 &   0, 1, 4, 6, 8, 9, 11, 13, 14, 17, 20, 21, 22, 24, 25, 29, 32, 34, 36, 38, 39, 40, 54, 56    & Poor\\
167 &   19, 46, 23    & Poor\\
168 &   59, 37, 6, 41, 47, 16, 52, 9, 26, 15    & Poor\\
169 &   9, 6    & Poor\\
170 &   0, 1, 4, 8, 11, 13, 14, 15, 17, 20, 21, 22, 24, 25, 27, 29, 30, 31, 32, 33, 34, 36, 37, 38, 39, 40, 51, 53, 54, 56, 57, 61, 64    & Poor\\
171 &   60, 37, 62, 61    & Poor\\
172 &   10, 4    & Poor\\
174 &   46, 19, 23, 60, 61, 62    & Poor\\
175 &   0, 1, 4, 8, 11, 13, 14, 17, 19, 20, 21, 22, 23, 24, 25, 29, 32, 34, 36, 38, 39, 40, 46, 54, 56    & Poor\\
176 &   0, 1, 4, 6, 8, 9, 11, 13, 14, 15, 17, 20, 21, 22, 24, 25, 27, 29, 30, 31, 32, 33, 34, 36, 37, 38, 39, 40, 51, 53, 54, 56, 57, 61, 64    & Poor\\
177 &   2, 5, 7, 8, 44, 46, 50, 19, 20, 23    & Poor\\
178 &   4, 6, 40, 9, 10, 61, 48, 49, 52, 54, 55, 59, 60, 42, 62, 63    & Poor\\
179 &   2, 5, 7, 9, 11, 14, 15, 16, 17, 22, 24, 26, 27, 30, 32, 34, 39, 43, 45, 47, 49, 57, 62    & Poor\\
180 &   2, 59, 4, 5, 7, 8, 41, 10, 44, 15, 16, 50, 52, 20, 26, 47    & Poor\\
\end{tabular}
\end{tiny}
  %  \end{landscape}
    %\clearpage% Flush page
}
%\newpage
	 
\afterpage{%
    %\clearpage% Flush earlier floats (otherwise order might not be correct)
    %\thispagestyle{empty}% empty page style (?)
    %\begin{landscape}% Landscape page
        \centering % Center table
        \begin{tiny}
\begin{tabular}{c|l|l}
181 &   60, 61, 62    & Poor\\
183 &   4, 5, 7, 10, 17, 20, 21, 26, 34, 36, 37, 38, 43, 44, 48, 54, 55, 56, 58, 59, 60, 61, 62, 63    & Poor\\
184 &   0, 1, 2, 3, 4, 5, 7, 8, 10, 19, 20, 24, 25, 27, 28, 29, 31, 32, 34, 35, 38, 39, 42, 44, 49, 50, 51, 55, 56    & Poor\\
186 &   0, 1, 4, 6, 8, 9, 11, 13, 14, 17, 20, 21, 22, 24, 25, 29, 32, 34, 36, 38, 39, 40, 54, 56    & Poor\\
187 &   0, 32, 2, 6, 5, 38, 7, 8, 12, 44, 14, 50, 20, 21, 9, 25, 33, 28, 31    & Poor\\
188 &   32, 51, 34, 35, 6, 38, 39, 9, 42, 49, 19, 56, 55, 24, 25, 27, 28, 29, 31    & Poor\\
189 &   0, 1, 4, 8, 11, 13, 14, 17, 20, 21, 22, 24, 25, 29, 32, 34, 36, 38, 39, 40, 54, 56    & Poor\\
190 &   0, 1, 3, 4, 5, 8, 10, 11, 13, 14, 17, 20, 21, 22, 24, 25, 29, 32, 34, 36, 38, 39, 40, 43, 44, 45, 46, 50, 54, 56, 57, 58, 63, 64    & Poor\\
191 &   0, 1, 4, 8, 11, 12, 13, 14, 17, 18, 19, 20, 21, 22, 23, 24, 25, 28, 29, 32, 34, 35, 36, 38, 39, 40, 46, 54, 56, 60    & Poor\\
192 &   4, 5, 7, 8, 10, 13, 15, 17, 18, 19, 20, 21, 22, 23, 26, 27, 29, 30, 31, 33, 34, 35, 36, 37, 38, 39, 41, 43, 44, 45, 46, 47, 48,  \\ &  50, 51, 53, 54,55, 56, 57, 58, 59, 61, 63, 64    & Poor\\
194 &   0, 1, 4, 6, 8, 9, 11, 13, 14, 17, 20, 21, 22, 24, 25, 29, 32, 34, 36, 38, 39, 40, 54, 56    & Poor\\
195 &   0, 1, 4, 8, 11, 13, 14, 17, 20, 21, 22, 24, 25, 29, 32, 34, 36, 38, 39, 40, 54, 56    & Poor\\
196 &   0, 1, 2, 3, 5, 6, 7, 9, 46, 19, 23, 60, 61, 62    & Poor\\
197 &   2, 35, 5, 6, 7, 8, 12, 44, 18, 60, 50, 20, 9, 28    & Poor\\
198 &   0, 1, 4, 8, 11, 13, 14, 17, 20, 21, 22, 24, 25, 29, 32, 34, 36, 37, 38, 39, 40, 54, 56    & Poor\\
199 &   8, 44, 50, 20    & Poor
\end{tabular}
\end{tiny}
    %\end{landscape}
    %\clearpage% Flush page
}

\end{document}